\documentclass{llncs}
\makeatletter
\def\@seccntformat#1{\@ifundefined{#1@cntformat}%
   {\csname the#1\endcsname\quad}  
   {\csname #1@cntformat\endcsname}
}
\let\oldappendix\appendix 
\renewcommand\appendix{%
    \oldappendix
    \newcommand{\section@cntformat}{\appendixname~\thesection\quad}
}
\makeatother
\usepackage{graphicx}
\usepackage{tabularx, booktabs}
\newcolumntype{Y}{>{\centering\arraybackslash}X}
\usepackage[figuresright]{rotating}
\usepackage[mathscr]{euscript}
\usepackage{placeins}

\begin{document}

\title{The Message or the Messenger? \\ Inferring Virality and Diffusion Structure from Online Petition Signature Data}
\titlerunning{petition virality}  
%
\author{Chi Ling Chan\inst{1} (ORCID 0000-0003-4665-714X), Justin Lai\inst{1} (ORCID 0000-0002-2591-8673), Bryan Hooi\inst{2} (ORCID 0000-0002-5645-1754), \and Todd Davies\inst{1} (ORCID 0000-0001-9082-4887)}

\institute{Stanford University, Stanford CA, USA\\
\email{callmechiling@gmail.com\\
jzlai@stanford.edu\\
davies@stanford.edu}
\and
Carnegie Mellon University, Pittsburgh PA, USA\\
\email{bhooi@andrew.cmu.edu}}
\authorrunning{Chan et al.} 
%
\tocauthor{Chi-Ling Chan, Justin Lai, Bryan Hooi, and Todd Davies}
%

\maketitle              

\begin{abstract}
Goel et al. \cite{goel} examined diffusion data from Twitter to conclude that online petitions are shared more virally than other types of content. Their definition of structural virality, which measures the extent to which diffusion follows a broadcast model or is spread person to person (virally), depends on knowing the topology of the diffusion cascade. But often the diffusion structure cannot be observed directly. We examined time-stamped signature data from the Obama White House's We the People petition platform. We developed measures based on temporal dynamics that, we argue, can be used to infer diffusion structure as well as the more intrinsic notion of virality sometimes known as infectiousness. These measures indicate that successful petitions are likely to be higher in both intrinsic and structural virality than unsuccessful petitions are. We also investigate threshold effects on petition signing that challenge simple contagion models, and report simulations for a theoretical model that are consistent with our data.
\end{abstract}
\section{Introduction}
This study infers the ``virality" and diffusion structure of petitions by examining the temporal dynamics of petition signatures. Viral characteristics can be understood as the opposite of broadcast characteristics. Whereas a \textit{broadcast} structure refers to large diffusion events in which a single source spreads content to a large number of people, \textit{viral} diffusion refers to a cascade of sharing events each between a sender and their associates. Intuitively, a petition that exhibits virality is more likely to attract signatures, in part, through the intrinsic appeal of its \textit{message}, whereas a petition that exhibits more broadcast characteristics is dependent on mass distribution by one or more well-connected senders (the \textit{messenger(s)}) in order to gain signatures.

We used time- and location-coded signature data from the Obama White House's We The People (WTP) petition site.\footnote{We the People petitions from the Obama years are archived at \url{https://petitions.obamawhitehouse.archives.gov}.} Whereas MoveOn.org and Change.org provide ``continuous user engagement'' through social media and email, WTP simply provides a static page for each petition, with standard sharing buttons. Since petitions on WTP are less likely to be broadcast, at least directly, than on other petition sites, we reasoned that they would be more likely to depend on person-to-person sharing, and hence viral characteristics, to reach the signature threshold for success \cite{karpf}. We wished to infer how petitions were shared without direct access to diffusion data. The data we did have -- about signatures  and the success or failure of each petition, constitute the variables of greatest interest for petitions, and, as is usually the case, the true diffusion data were not observable because they involve an unknown amount of private communication (including emails, phone calls, and face to face dialogue). We therefore sought indirect ways to measure viral and broadcast components of diffusion.  
 
%
\section{Related Work}
\subsubsection{Virality}

Recent literature has examined the virality of online petitions and their differences from other social media. Compared to other online activities, petition-signing typically requires personal endorsement/commitment before sharing. As such, researchers have noted that Twitter cascades about petitions exhibit more \textit{structural virality} than those for news, pictures, and videos \cite{goel,goel2}. Structural virality (SV) is a continuous measure that distinguishes between a single, large broadcast (high SV) and viral spread over multiple generations (low SV). 

Structural virality is defined as the average distance between all pairs of nodes in a diffusion tree, a quantity that, for a given cascade size (number of Tweets and Re-Tweets) is minimized when all Re-Tweets are direct offspring of a single source Tweet (pure broadcast diffusion). It is maximized when each Re-Tweet is itself directly Re-Tweeted just once, indicating a string of successors influenced by each predecessor, rather than one predecessor with a great deal of influence. Goel et. al found little to no correlation between structural virality and popularity (cascade size) within any given type of shared content, including petitions (for which they found $r=$.04) \cite{goel}.

\subsubsection{Temporal Adoption Patterns}

\begin{figure}[h]
\centering  
\includegraphics[width=0.8\textwidth]{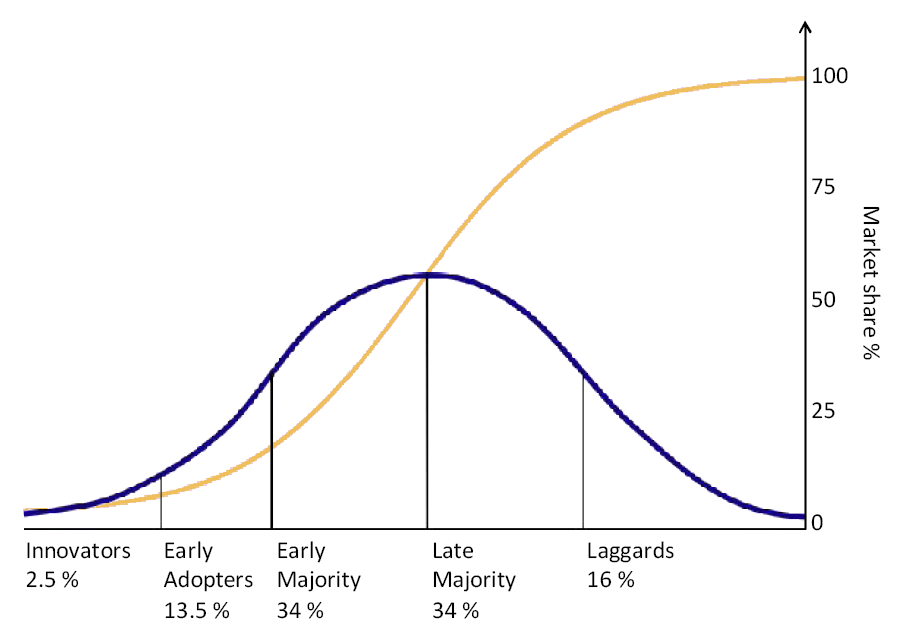}
\caption{The classical adoption pattern, showing different phases of adoption for a cumulative S-shaped curve. (Graph from Wikimedia Commons, \url{https://commons.wikimedia.org/wiki/File:Diffusionofideas.PNG}, Public Domain.)} \label{classical_adoption_curve}
\end{figure}

In the literature on diffusion and contagion, the dominant model posits an S-shaped cumulative adoption curve \cite{coleman,bass,rogers,young,valente,iyengar}, which exhibits an initial period of exponential growth that levels off when the population runs out of potential adopters (see Figure \ref{classical_adoption_curve}). 

Whether this diffusion pattern applies in the online setting has been a question of interest, given the increasing prominence of online mobilization. Investigations of online network mobilizations have identified an S-shaped curve, with critical mass reached only after participants have responded to evidence from early participants \cite{gonz}. However, evidence remains mixed. In a study that tracks the growth curves of 20,000 petitions on the government petitioning site in the United Kingdom, Yasseri et al. \cite{yasseri} noted that a petition's fate is virtually set after the first 24 hours of its introduction, a finding echoed in a study \cite{hale2} which tracked 8000 petitions on the UK petitioning site No. 10 Downing Street. The S-shaped period of stasis before reaching a critical point appears largely absent, consistent with most studies of online petitioning. These findings call into question the explanatory power of the S-curve model for online petitions. Given that research on online petitions is a relatively new area of study \cite{wright}, however, there remains a lack of empirical calibration and external validity - a point acknowledged by authors of most of these studies \cite{gonz,hale2}.

A second question has to do with the diffusion mechanism behind observed patterns. Empirical studies have largely interpreted the S-curve as evidence of social contagion \cite{rogers,bass},  but others suggest that the same curve could arise from broadcast distribution mechanisms such as mass media sharing \cite{vandenbulte}. It remains ambiguous whether viral diffusion is in fact the diffusion mechanism driving growth momentum, as is typically assumed in classical diffusion studies.  Related research has shown that information cascades in online networks occur rarely, and studies of online petitions have found that the vast majority of signatures are dominated by a tiny fraction of massively successful petitions \cite{yasseri,jung,hale}. Events that `go viral' are an exception rather than the rule \cite{goel}. Untangling broadcast from viral diffusion mechanisms, however, has been difficult largely because these studies have based themselves on aggregate diffusion data.

Recent studies have been able to overcome this limitation by analyzing cascade structures directly. Goel et al. analyzed cascades from a billion diffusion events on Twitter and offered a fine-grained analysis of how viral and broadcast diffusion interact \cite{goel}. They found that large diffusion events exhibit extreme diversity of structural forms, and demonstrate various mixes of viral and broadcast diffusion, such that the S-curve is but one of many combinations. Extending Goel's analysis of viral and broadcast diffusion, we might ask related questions about how patterns develop over time: When do viral and/or broadcast diffusion set in, and how do they combine to generate the adoption patterns observed? Hale et al. observed that signatures are typically gathered via ``punctuated equilibria,'' specific points that trigger large cascades within a short time, resulting in leptokurtic distributions (characterized by sharp peaks in signature counts) \cite{hale2}, which suggest broadcast events. 

\subsubsection{Goal Thresholds} 

A related but less well-studied aspect of online petitioning and diffusion has to do with the effects of threshold requirements that define petition success. Signature thresholds are a typical feature in most online petitions and are crucial for goal-setting by campaigners. Social psychology studies demonstrate that people invest greater effort as they approach a goal \cite{kivetz,cheema,koo}, a phenomenon known as the goal-gradient hypothesis, first described by behaviorist Clark Hull (1934) observing rats running faster as they approach a food reward in a maze \cite{hull}. For online petitions, there is suggestive evidence in Hale (2013) for threshold effects at the 500-signature mark (the minimum required for an official response on No.10 Downing Street)\cite{hale}, but it remains unclear if this can be extrapolated across other petition platforms such as WTP. \\
\section{Data and Methods}
\subsection{WTP petitions data}

In this observational study, we relied on an aggregated adoption dataset of 3682 publicly searchable petitions using the public API of WTP.gov, the official online petitioning platform of the White House.  These petitions were created within a time period that began from the inception of the platform on September 20, 2011 and ended on March 30, 2015, as our study commenced.

Two critical thresholds for We the People should be noted. First, to be publicly listed and searchable within the site, a petition had to reach 150 signatures within 30 days. Second, to cross the second threshold for review by the White House and be distributed to policy officials for an official response, a petition had to obtain 100,000 signatures (25,000 until January 2013) within 30 days. Responses were posted and linked to the petition on WhiteHouse.gov, and emailed to all petition signers. All petitions in our dataset crossed the 150 signatures threshold, and were publicly listed (the API allowed the retrieval of only petitions that were publicly searchable). 

\subsection{Variables in signatures and petitions dataset}

The WTP API provided data on both petitions and their individual signatures submitted via the petition site. Data on petitions included timestamp of creation, the body of text that campaigners submitted, status at the time of the API query (open/pending response/responded/closed), and the signature count. The data included the time a signature was submitted, as well as geographical details (state, Zipcode if the signature was from the U.S., country, city) of the signatory.

From the API, we assembled a dataset comprising all signatures and petitions and organized them into the following variables: 1) petition ID; 2) signature ID; 3) Unix timestamp of signature; and 4) Zipcode of signatory. This was then merged with a dataset of petitions containing the following data: 1) petition ID; 2) petition title; 3) petition description; 4) signature count; 5) signature status; and 6) Unix timestamp of creation (Unix timestamps were recoded to reflect number of days since a petition's creation).\footnote{Our data are available at \url{https://github.com/justinlai/petitiondata}.}

For this study, a petition was considered successful if it reached the 100,000 (or 25,000) signature threshold necessary for White House review.\footnote{For an alternative perspective on e-petition ``success,'' see Wright (2016) \cite{wright2}.} Using this criterion, a large majority (98.4\%) of petitions failed. Of all visible petitions, 1.6\% reached the 100,000 threshold. This success rate is consistent with the predicted pattern that only a small fraction of campaigns eventually succeed \cite{yasseri}. 

\subsection{Some measures of virality}
\label{some_measures_of_virality}
Because we could not observe the diffusion network of a petition on WTP, we could not use the same measure of structural virality (SV) used by Goel et al. \cite{goel}. Furthermore, as they note, their concept of SV may have no relationship to the concept of ``infectiousness,'' or the probability that, in this case, a recipient of a petition announcement will sign the petition.\footnote{Indeed, for some other types of content such as the spread of memes, initial infectiousness appears not to be a good predictor of later success \cite{salganik,weng}.} Therefore, we propose indirect measures of SV, which we also distinguish from infectiousness which we call \textit{intrinsic virality} (IV). An indirect measure of SV is the \textit{exceed ratio}, while IV may be assessed through the \textit{first-day, second-day signature comparison}.

\subsubsection{Exceed Ratios}
The \textit{exceed ratio} is a measure indicating the contribution of temporal peaks to a petition's total signature count. A temporal peak is a period (e.g. day or hour) when the signature count exceeds those within both of its adjacent time periods of the same duration. For every temporal peak, we calculate the signatures received on that peak day minus the number received on either the day before or the day after, whichever is larger. The total exceed ratio $E_{Tot}$ is the sum of these differences across all temporal peaks, divided by the total signature count. Notationally, $E_{Tot}$, for a given petition over $T$ time periods, in which $S[i]$ signatures are obtained in period $i$, and $L$ refers to the set of all peak periods within $T$, is thus defined as: 

$$E_{Tot} = \frac{\Sigma_{i\in L} (S(i) - \max[S(i-1), S(i+1)]) }{\Sigma_{i=1}^{T} S(i)} $$

The total exceed ratio is a measure of broadcast-ness, and therefore are inverse measure of structural virality. Broadcast content is more likely to rely on large diffusion events in which a period's signature count is larger than those in its adjacent periods, whereas viral content would likely have fewer such events.

We can also calculate a global-peak-only exceed ratio $E_{GPO}$ by dividing the adjacent-periods signature difference for just the global peak period by total signatures, as an indication of the effect of the largest broadcast event. 
%
%





\subsubsection{First-Day, Second-Day Signature Comparison}

The \textit{first-day, second-day signature comparison} (FDSD) for a given petition asks whether the number of signatures received on the first day is exceeded by those on the second day. We posit this as a simple measure of intrinsic virality (IV, or infectiousness, viz, intrinsic message appeal for signing and sharing), on the assumption that a petition with high IV will be more likely both to be passed on and to be signed by recipients than will one with lower IV. The first day is the best day to make this follow-on comparison if we assume petitions are most likely to be announced in broadcast events on their first day. If this is the case, then FDSD provides the best standard comparison across petitions, which on other days are likely to vary in whether or not they are broadcast. Since many petitions gain no traction after the first day, the FDSD also allows us to capture that lack of enthusiasm for the largest number of petitions under study.
\section{Results}
\subsection {Overall adoption patterns} 
Overall, a total of 24.5 million signatures were collected by 3682 petitions in our dataset, of which 59 (1.6\%) reached the 100,000 signature threshold required for a response from the White House. Successful petitions garnered 31.8\% of total signatures.

Each petition's signature data can be plotted as an adoption curve, showing the number of signatures reached day by day. Figures \ref{subset_success} and \ref{subset_fail} in Appendix A show temporal signature histograms for randomly chosen successful and unsuccessful petitions, to give a sense of how these look. Figures \ref{threshold_effect} and \ref{adoption_all_cdf_color} of Appendix B show aggregated cumulative adoption patterns and the 30 day temporal thresholds for all the petitions (successful and unsuccessful).

\subsection{Structural virality vs. broadcast events}

We found that unsuccessful petitions have a\textbf{ 47.4\%} higher average total exceed ratio $E_{Tot}$ than successful petitions in the daily distribution of signatures. For the hourly distribution of signatures, this rises to \textbf{55.4\%} ($p<$.0001 for both comparisons by a two-tailed t-test). This suggests that successful petitions are less dependent on broadcast events (peaks) for growth, and therefore higher in structural virality.

\begin{table}[h]
\centering
\caption{Average total exceed ratio $E_{Tot}$ for all petitions: successful versus unsuccessful.}
\label{tab:exceed_ratios}
\def\arraystretch{1.5}%
 \begin{tabularx}{0.85\textwidth}{ | X | X | X | } 
 \hline
 &  Successful (N=59) &  Unsuccessful (N=3623) \\ [0.5ex] 
 \hline
 Daily & 0.152 (sd=.13) & 0.224 (sd=.04) \\ 
 Hourly & 0.148 (sd=.09)& 0.230 (sd=.03) \\ 
 \hline
\end{tabularx}
\end{table}

The daily global-peak-only exceed ratio $E_{GPO}$ for successful petitions was 0.105 (sd=.11), and for unsuccessful ones was 0.155 (sd=.19). This appears to conflict with the statement by Goel et al. that ``if popularity is consistently related to any one feature, it is the size of the largest broadcast,'' since in our data the indicator of a larger single broadcast is higher for unsuccessful (therefore less popular) petitions ($p=$.042 by a two-tailed t-test).\footnote{In footnote 10 on p. 187, Goel et al. clarify that this statement applies to normalized and not just to absolute size \cite{goel}.} 

\subsection{Intrinsic virality (infectiousness)}

Among the 59 successful petitions, 68\% had more signatures in their second day than their first day, but among the 3623 unsuccessful petitions, this percentage was only 38\%. Thus, there is a clear relationship between a petition's success and having more signatures in its second than its first day ($\chi^2$ test $p$-value $<10^{-5}$).

This finding agrees with our earlier intuition that poorly performing petitions tend to receive an initial burst of signatures but then decay quickly. Moreover, this measure is particularly interesting as it only relies on the first two days, and thus can be used as an early indicator of whether a petition is likely to succeed. By the reasoning in subsection \ref{some_measures_of_virality}, we infer that successful petitions are higher in intrinsic virality than are unsuccessful ones. This may seem like an obvious truth, but it contradicts influential models of diffusion which imply that ``the largest and most viral cascades are not inherently better than those that fail to gain traction, but are simply more fortunate'' \cite{goel} (citing \cite{watts}).


\subsection{Additional measures: viral and broadcast diffusion across all petitions}
\label{additional_measures}

We can also analyze our dataset as a whole, without distinguishing between petitions that do and do not pass the success threshold. Additional measures of interest for looking at virality across all petitions are described in Table \ref{tab:shape_measures}. A petition's `global peak' is the day during which it received the most signatures (where day 1 is the day the petition was introduced). This is another indirect way to measure intrinsic virality, on the assumption that a petition with more appeal will be more likely to grow in its signature count rather than die out, and hence would be expected to have a later global peak. The dependent variable, $total$, is the total number of signatures a petition acquires over the 60 day period. Here we will refer to more and less popular petitions to indicate numerical differences rather than the binary, successful and unsuccessful categories.

\begin{table}[h!]
\centering
\caption{Measures of shape examined in this section (with the type of virality measured noted in brackets)}
\label{tab:shape_measures}
\begin{tabular}{|r|l|}
\hline
{\bf Measure of Shape}        & {\bf Interpretation}                                                                                                             \\ \hline
{\it Skewness}                & \begin{tabular}[c]{@{}l@{}}whether distribution has larger \\ `tails' extending to right (positive) [IV] \end{tabular}                \\ \hline
{\it Kurtosis}                & how peaked a distribution is [SV]                                                                                                   \\ \hline
{\it Location of global peak} & \begin{tabular}[c]{@{}l@{}}the day on which the petition received\\ the most signatures [IV] \end{tabular}                             \\ \hline
{\it Number of local peaks}   & \begin{tabular}[c]{@{}l@{}}number of days on which the petition\\ received more signatures than on \\ adjacent days [SV]  \end{tabular} \\ \hline
\end{tabular}
\end{table}

\subsubsection{Regression Results}
We perform linear regressions of the number of signatures over our measures in Table \ref{tab:shape_measures}. The results are shown in Table \ref{tab:peak_regression}.

\begin{sidewaystable}
\centering 
  \caption{Linear regression shows a highly significant positive relationship between peak\_day and total signatures (total). Model 3 shows that this relationship remains significant when controlling for the number of peaks, variance, skewness, and kurtosis. Model 4 shows that this remains significant when log transforming the dependent variable (number of signatures).} 
  \label{tab:peak_regression} 
\scriptsize 
\begin{tabular}{@{\extracolsep{5pt}}lcccc} 
\\[-1.8ex]\hline 
\hline \\[-1.8ex] 
 & \multicolumn{4}{c}{\textit{Dependent variable:}} \\ 
\cline{2-5} 
\\[-1.8ex] & \multicolumn{3}{c}{total signatures} & log(total) \\ 
\\[-1.8ex] & (Model 1) & (Model 2) & (Model 3) & (Model 4)\\ 
\hline \\[-1.8ex] 
 skewness & 5,009.992$^{***}$ &  & 5,875.782$^{***}$ & 0.558$^{***}$ \\ 
  & p = 0.00003 &  & p = 0.00000 & p = 0.000 \\ 
  & & & & \\ 
 kurtosis & $-$585.658$^{***}$ &  & $-$532.246$^{***}$ & $-$0.062$^{***}$ \\ 
  & p = 0.00001 &  & p = 0.00003 & p = 0.000 \\ 
  & & & & \\ 
 global\_peak\_day &  & 262.890$^{***}$ & 227.812$^{***}$ & 0.008$^{**}$ \\ 
  &  & p = 0.000 & p = 0.00000 & p = 0.013 \\ 
  & & & & \\ 
 num\_local\_peaks &  &  & 2,103.878$^{***}$ & 0.150$^{***}$ \\ 
  &  &  & p = 0.000 & p = 0.000 \\ 
  & & & & \\ 
 Constant & $-$342.411 & 4,985.854$^{***}$ & $-$22,660.150$^{***}$ & 5.455$^{***}$ \\ 
  & p = 0.860 & p = 0.000 & p = 0.000 & p = 0.000 \\ 
  & & & & \\ 
\hline \\[-1.8ex] 
Observations & 3,682 & 3,682 & 3,682 & 3,682 \\ 
R$^{2}$ & 0.006 & 0.010 & 0.087 & 0.089 \\ 
Adjusted R$^{2}$ & 0.005 & 0.010 & 0.086 & 0.088 \\ 
Residual Std. Error & 19,316.340 (df = 3679) & 19,271.360 (df = 3680) & 18,518.850 (df = 3677) & 1.364 (df = 3677) \\ 
F Statistic & 10.355$^{***}$ (df = 2; 3679) & 36.997$^{***}$ (df = 1; 3680) & 87.054$^{***}$ (df = 4; 3677) & 90.216$^{***}$ (df = 4; 3677) \\ 
\hline 
\hline \\[-1.8ex] 
\textit{Note:}  & \multicolumn{4}{r}{$^{*}$p$<$0.1; $^{**}$p$<$0.05; $^{***}$p$<$0.01} \\ 
\end{tabular} 

\end{sidewaystable}

As shown in Columns $1$, $3$ and $4$ of Table \ref{tab:peak_regression}, we find that skewness and kurtosis are also significantly correlated with signature count. Under all three model specifications, petitions with right skewed distributions (i.e. larger right tails) tend to end up with more signatures (which indicates intrinsic virality, i.e. more signatures late in the process due to more people signing and passing on the petition), as do petitions with lower kurtosis (i.e. having less sharp peaks). This latter finding would be predicated by higher structural virality, since kurtosis is related to our exceed ratio as an indicator of events (the inverse of virality).

As shown in column $2$ of Table \ref{tab:peak_regression}, linear regression suggests that on average, a petition that peaks $1$ day later ends up with $262.89$ more signatures ($p \approx 1.3\times 10^{-9}$). The relationship remains about as strong, and still highly significant, when we control for the number of local peaks, and the skewness and kurtosis of the petition's temporal distribution (column $3$) as well as when we replace the dependent variable by its logarithm (to ensure that petitions with large signature counts do not excessively influence the fitted coefficients).

\begin{figure}[h!]
\centering  
\includegraphics[width=0.9\textwidth]{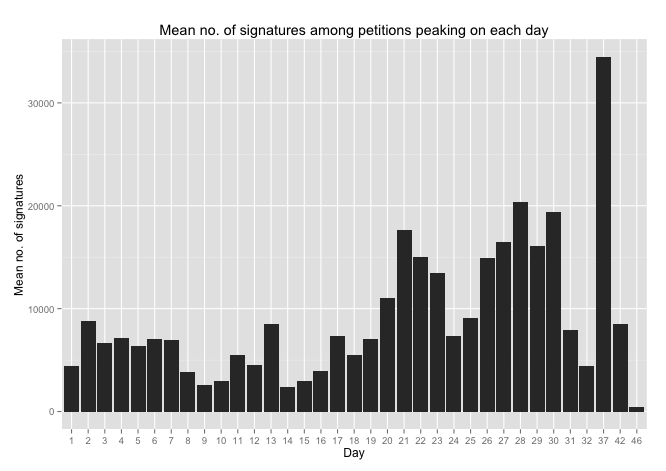}
\caption{The chart indicates the mean number of signatures of petitions with a global peak on a given day (with missing days after day 32 to a lack of instances). Petitions which peak on later days tend to end up with a higher total number of signatures than those which peak on earlier days.} \label{mean_signatures_by_peak_day}
\end{figure}

Hence, we find that petitions with later global peak days tend to end up with more signatures than petitions with earlier peaks, which we take to be an indication of intrinsic virality. This finding is also illustrated in Figure \ref{mean_signatures_by_peak_day}, in which petitions are separated into those with global peaks on day $1$, $2$, and so on; and we observe, the mean numbers of signatures seems to broadly increase as the peak gets later, which agrees with the regression findings.

As noted above, more popular petitions have more local peaks. However, further analysis reveals that this phenomenon occurs mainly because less popular petitions receive very few signatures after day $30$ (as discussed in our threshold section) particularly when they do not reach the $100,000$ mark, and hence have fewer local peaks. Indeed, when we consider only days $1$ to $30$ and perform another regression of $\log(total)$ against $num\_local\_peaks$, the regression coefficient becomes much smaller and no longer statistically significant (coefficient of $-0.009, p=0.437$).\\

\subsection{Threshold Effects}

Our analysis of petition adoption curves reveals: (a) a goal-gradient threshold effect, i.e. petitions start to receive fewer signatures after reaching the $100,000$ signatures mark; and (b) a temporal threshold effect, i.e. if a petition becomes $30$ days old without receiving $100,000$ signatures, it suddenly starts to receive fewer signatures. (See Figs. \ref{threshold_effect} and \ref{adoption_all_cdf_color} in Appendix B.) We take this as evidence that WTP site users are paying attention to the social context when they decide to sign, which is at odds with a ``simple contagion'' model in which the probability of signing as a result of each new communication remains the same (see, e.g. \cite{centola,goel,grano}). 

\section{A Theoretical Model}
In this section we describe a model that explains the observed signatures as a mixture of broadcasts (which induce a group of users to sign the petition, e.g. a news broadcast) and viral spreading of the petition from people who have just signed it to new users. Under this model, at each time step, a petition has a small probability of a broadcast occurring. Each broadcast brings in a number of users, where the number is drawn from a log-normal distribution, which allows for large variance in broadcast sizes similar to what we observe in actual data. Naturally, this distribution can be replaced by any appropriate distribution in other applications depending on the researcher's prior beliefs. 

At the same time, viral spread is happening constantly - each user who has just signed the petition has a small probability of spreading the petition to each other user who has not signed it yet. The strength of the viral spread can be parametrized by the basic reproduction number $R_0$, the average number of people that each signer spreads the petition to in a completely susceptible population. In addition, since we observe in the real data a fairly constant and low `background' level at which users sign the petition, we also add a similar low background probability for each user in the population to sign the petition at each time step, independent of the existing broadcast and viral mechanism. 

For our simulations, the probability of broadcasts was chosen to give an average of $3$ broadcasts per petition, with broadcast size following a log-normal distribution: if $X$ is a broadcast size, then $\log X \sim \mathcal{N}(\mu, \sigma^2)$, where we use $\mu=5, \sigma=1.5$. There is always at least one broadcast, and the first broadcast occurs on day $1$. For viral spread, the initial susceptible population is set at $10000$, and $R_0$ is chosen from a uniform distribution between  $0.7$ and $1.9$. The background level is set such that each user who has not signed the petition has a $0.002$ chance of signing it at each time step. 

In this model, $R_0$ could be thought of as what we have called ``instrinsic virality'' (IV) of the \textit{message}, in that it varies across petitions and measures the likelihood of signing and passing along that petition across all recipients in the population. The average broadcast size, by contrast, is assumed to be the same for each petition, since we assume that to be a feature of the population (the \textit{messengers}) rather than of the petition itself. 


\subsubsection{Replicating Empirical Findings on Shape and Success} \label{sec:replicating}

\begin{table}[h!] \centering 
  \caption{Comparison between regression coefficients using simulated data (left col.) vs. actual data (right col.). Regression coefficients are all significant and match by sign; however, there is a stronger effect of num\_local\_peaks for the actual data. Variables: global\_peak\_day: which day the most signatures were received. num\_local\_peaks: number of days at which more signatures were received than the previous and next day.} 
  \label{tab:sim_vs_actual} 
\footnotesize 
\begin{tabular}{@{\extracolsep{5pt}}lcc} 
\\[-1.8ex]\hline 
\hline \\[-1.8ex] 
 & \multicolumn{2}{c}{\textit{Dependent variable:}} \\ 
\cline{2-3} 
\\[-1.8ex] & \multicolumn{2}{c}{log(total)} \\ 
 & Simulated & Actual Data \\ 
\\[-1.8ex] & (1) & (2)\\ 
\hline \\[-1.8ex] 
 global\_peak\_day & 0.007$^{***}$ & 0.008$^{**}$ \\ 
  & p = 0.000 & p = 0.013 \\ 
  & & \\ 
 num\_local\_peaks & 0.024$^{***}$ & 0.150$^{***}$ \\ 
  & p = 0.000 & p = 0.000 \\ 
  & & \\ 
 skewness & 0.453$^{***}$ & 0.558$^{***}$ \\ 
  & p = 0.000 & p = 0.000 \\ 
  & & \\ 
 kurtosis & $-$0.028$^{***}$ & $-$0.062$^{***}$ \\ 
  & p = 0.000 & p = 0.000 \\ 
  & & \\ 
 Constant & 5.991$^{***}$ & 5.455$^{***}$ \\ 
  & p = 0.000 & p = 0.000 \\ 
  & & \\ 
\hline \\[-1.8ex] 
Observations & 5,000 & 3,682 \\ 
R$^{2}$ & 0.298 & 0.089 \\ 
Adjusted R$^{2}$ & 0.298 & 0.088 \\ 
Residual Std. Error & 0.403 (df = 4995) & 1.364 (df = 3677) \\ 
F Statistic & 530.431$^{***}$ (df = 4; 4995) & 90.216$^{***}$ (df = 4; 3677) \\ 
\hline 
\hline \\[-1.8ex] 
\textit{Note:}  & \multicolumn{2}{r}{$^{*}$p$<$0.1; $^{**}$p$<$0.05; $^{***}$p$<$0.01} \\ 
\end{tabular} 
\end{table} 

Are the empirical findings from subsection \ref{additional_measures} relating the shape of a petition's adoption curve and its popularity also present in the simulations? If they are, this provides a possible explanation for the empirical findings; if not, they suggest a way of improving the model. To answer this question, we simulate $5000$ petitions using the broadcast and viral model, and do a linear regression of the logarithm of the total number of signatures received by a petition against measures of shape as we did earlier (in Table \ref{tab:peak_regression}). As before, we use the logarithm of the number of signatures as the response variable to prevent outliers from excessively influencing the fit. Table \ref{tab:sim_vs_actual} shows the regression coefficients when using these simulations (col. 1) compared to the original regression coefficients for the actual data (col. 2). Table \ref{tab:sim_vs_actual} shows that all the coefficients for the regression on simulated data are significant with the same sign as in the original petition data. Most are of fairly similar magnitudes, with the exception of num\_local\_peaks, which has a stronger effect in the actual data. However, we observed earlier that this variable is significant in the actual data largely as an artifact of the long runs of zeros for less successful petitions.

Since the simulations are based on a simple model, we can explain these regression findings. Petitions with low $R_0$ peak early when they receive an initial broadcast but then lose momentum extremely quickly due to the lack of strong viral spread; hence, they have earlier global peaks, short right tails, and a highly peaked distribution. Petitions with high $R_0$ accumulate signatures more gradually due to having stronger viral spread, then lose momentum gradually as the population runs out of users who have not signed the petition, thus having later global peaks, larger right tails, and a less peaked distribution. Since petitions with high $R_0$ end up with more total signatures, these account for the regression coefficients.

This does not necessarily imply that the same effects are present in the actual data. But these simulations provide a plausible explanation for the empirical findings that more successful petitions have later global peaks, more skewed and less peaked distributions. The simulation results suggest that under a simple model of viral spread, as long as different petitions have varying values of $R_0$ (i.e. rate of viral spread), we should expect correlations between total number of signatures and these measures of petition shape. As we have observed in our simulations, this is because higher $R_0$ petitions have different characteristic shapes than low $R_0$ petitions, and also end up with more signatures.

\section{Discussion and Future Work}
We have studied the temporal dynamics of adoption and diffusion patterns in online petition-signing, in order to understand what makes petitions gain traction and growth momentum. In this final section, we return to the questions that motivated this study and discuss theoretical and practical implications of our observations and modeling. 

While Goel et. al noted that there is a very weak correlation for popularity and structural virality (the average distance between nodes) for petition sharing on Twitter \cite{goel}, our research finds that our measures of intrinsic virality (first-day second-day comparison, skewness, and global peak day) are highly predictive of petition popularity/success. Our measures of exceed ratio and kurtosis indicate further that threshold-successful and/or more popular petitions are higher in structural virality. Intrinsic virality we take to be a property of a petition's message, whereas structural virality is the inverse of broadcast-ness, which we take to be a feature of its messengers.

The fact that intrinsic virality, which is based on the appeal of the message rather than how it is spread, appears to predict success for petitions on We the People, is a partial answer to a question posed by Goel et al. about the relationship between structural virality and what they call ``infectiousness" (intrinsic virality). They look at different models in which infectiousness is assumed to be either fixed or varying between different messages (petitions, in our case), and remain uncommitted about which one better describes real data. Our results for the first-day, second-day comparison and other intrinsic virality measures argue that not all messages are created equal, and that early indications of a message's intrinsic appeal, before the diffusion structure has had a chance to be determined, are correlated with eventual success for petitions. Two other findings challenging whether the assumptions of Goel et al. apply to WTP are that (a) the global-peak-only exceed ratio for daily signature totals was lower for successful than for failed petitions on We the People, which indicates that the largest (relative) broadcast event may not be the best predictor of petition popularity/success; and (b) signatures on WTP exhibit a strong threshold effect (consistent with the goal gradient hypothesis), which is at odds with a simple contagion model, of which theirs is a special case (\cite{goel}, p. 189, footnote 14).

Regression analysis over all the petitions, in subsection \ref{additional_measures}, indicated that more successful petitions  exhibit: 1) a later global peak; 2) a larger right tail (positive skewness); 3) a less peaked distribution (lower kurtosis). Features 1 and 2 are indicative of intrinsic virality (IV), while feature 3 is indicative of structural virality (SV). Petitions that under-perform often experience early bursts of momentum at the outset, but the decay of such spikes is usually rapid. Based on our simulations, we find that a simple model combining broadcasts and viral diffusion, in which different petitions have different strength of viral diffusion (or $R_0$) (analogous to IV) can account for these three findings, primarily due to the different characteristic shapes between high and low $R_0$ petitions. 

Our FDSD variable finding revealed that a petition is likely to fail if the number of signatures gathered on its second day is lower than its first day, indicating it has low intrinsic virality. But previous research has shown the first day alone to be a very important predictor of petition success \cite{hale2,yasseri}. This could be taken as evidence for the importance of the initial broadcast event, in addition to structural and intrinsic virality, for petition popularity. Absent an effective broadcast, it is highly unlikely that a viral effect will set in and bring about the necessary momentum for growth in support. 

The WTP data contain location- as well as timestamped signatures. This opens the possibility of further testing for geographical diffusion effects. An initial attempt to predict petition success from average land distance between Zipcodes in adjacent signature pairs did not uncover a difference between successful and unsuccessful petitions. However, this may be  due to the fact that the overall average distance confounds both diffusion distance between signers (which might be lower for viral transmission) and the the fact that more popular petitions are likely eventually to find an audience across larger distances than are less popular ones that fizzle early. This question awaits a good measure that can disentangle these potentially opposing effects.

\section{Acknowledgements}

We wish to thank Marek Hlavac for technical assistance, and Lee Ross and Howard Rheingold for timely and valuable feedback on an earlier version of this work (which was submitted by the first author as her masters thesis \cite{chan}), as well as three anonymous reviewers for their helpful comments. 

\newpage

\begin{appendix}
\section{Signature Graphs for Individual Petitions}
\begin{figure}[h!]
\centering
\includegraphics[width=\textwidth]{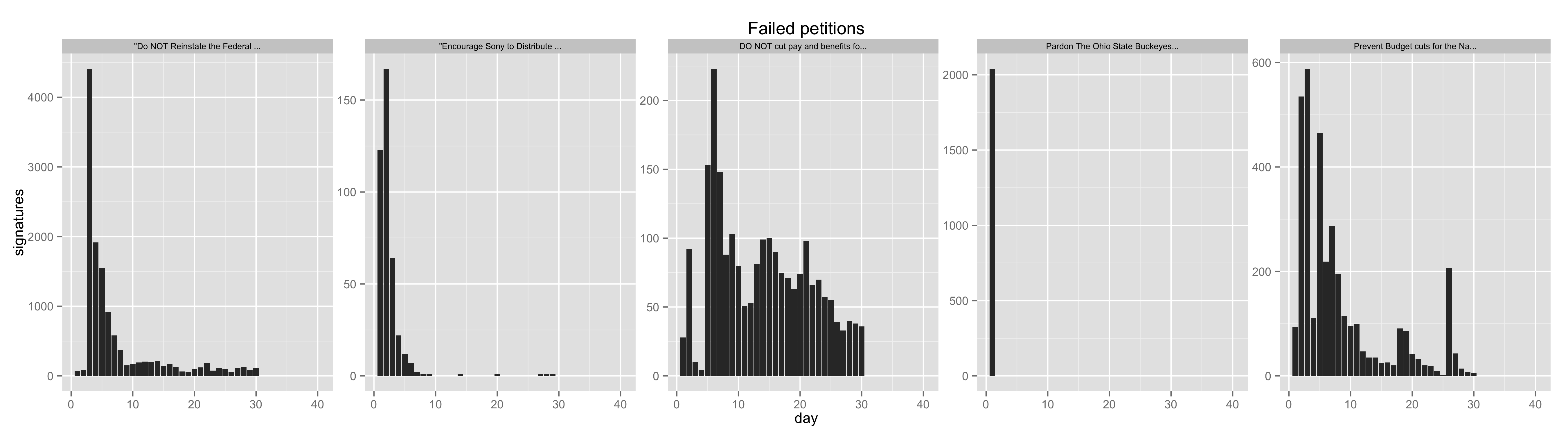}
\caption{Temporal distribution of $5$ randomly chosen petitions which did not reach the $100,000$ signature mark.} \label{subset_fail}
\vspace{60pt}
\centering
\includegraphics[width=\textwidth]{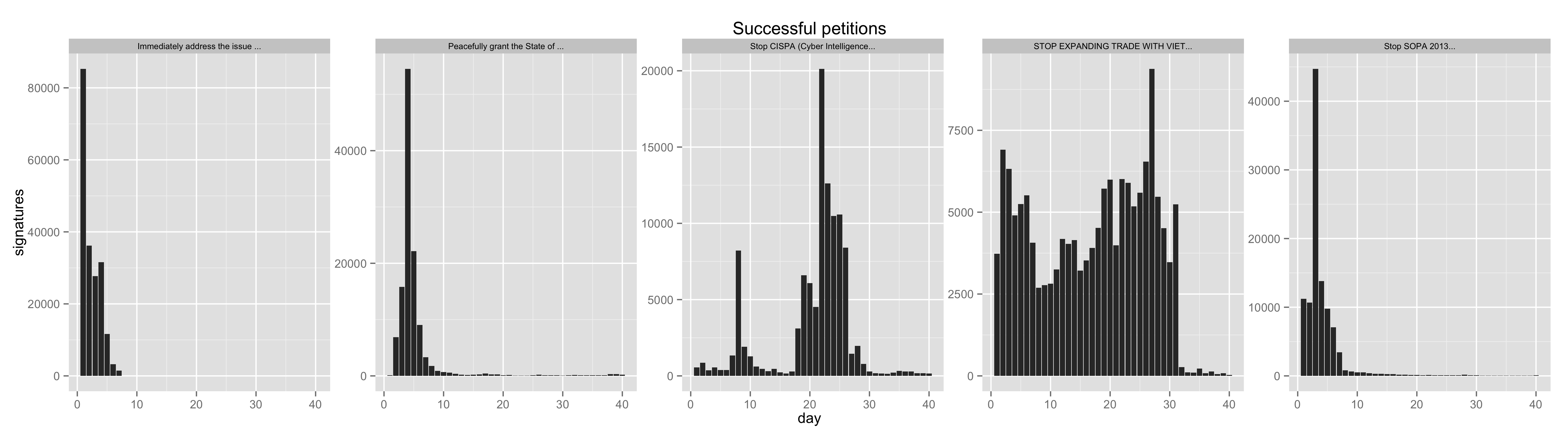}
\caption{Temporal distribution of $5$ randomly chosen petitions which succeeded in reaching the $100,000$ signature mark} \label{subset_success}
\end{figure} 
\newpage
\section{Aggregated Temporal Signature Graphs}
\begin{figure}[h!]
 \centering
  \includegraphics[width=0.65\textwidth]{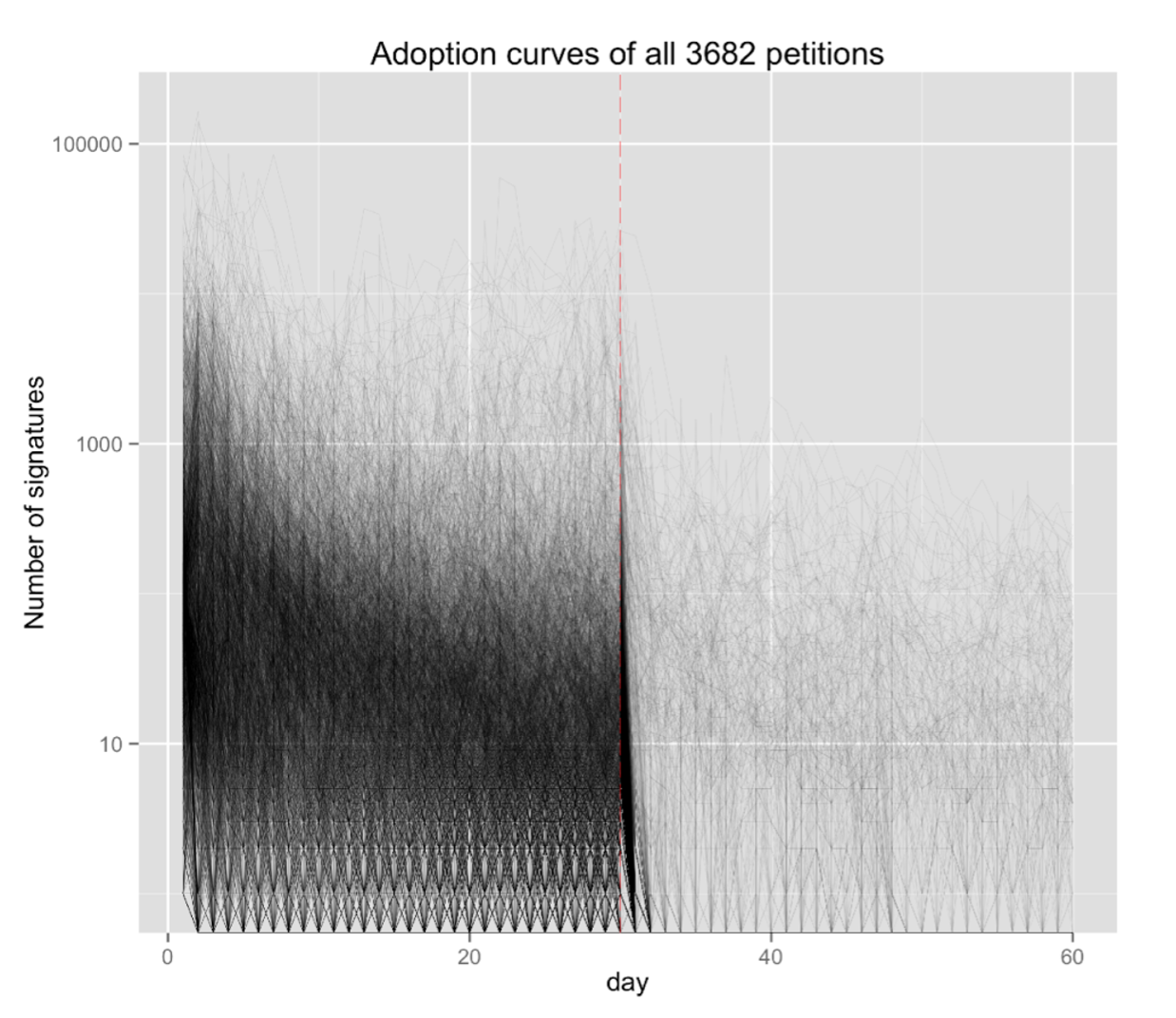}
  \caption{Adoption curves capturing daily accumulation of signatures in 3682 petitions across a 60-day period. A clear spike indicating a surge in support for a large number of petitions right before the 30-day deadline.}
  \label{threshold_effect}
  \centering
  \includegraphics[width=0.65\textwidth]{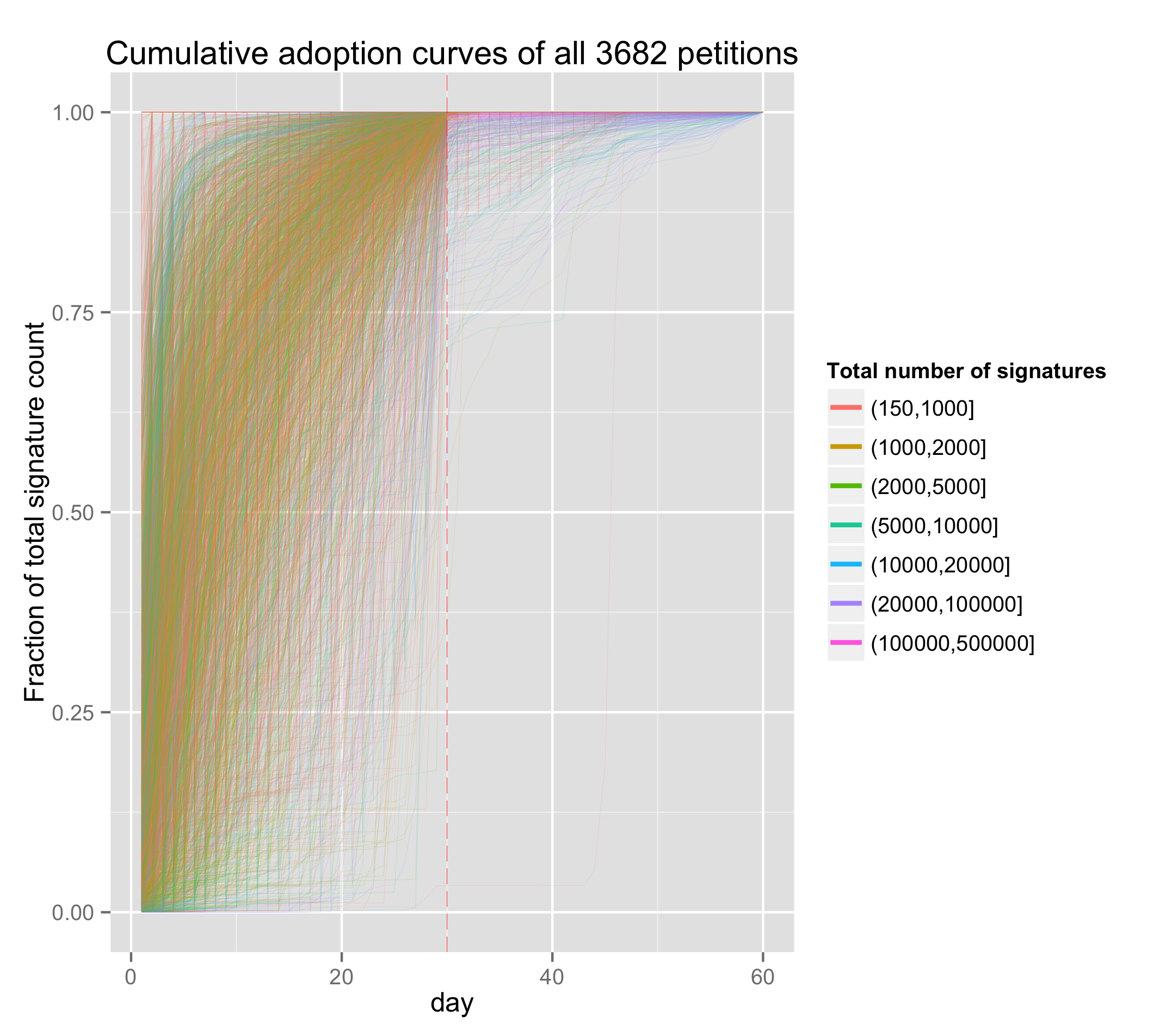}
  \caption{Adoption curves capturing daily accumulation of signatures in 3682 petitions across a 60-day period. A clear spike indicating a surge in support for a large number of petitions right before the 30-day deadline.}
  \label{adoption_all_cdf_color}
\end{figure}
\end{appendix}

\FloatBarrier

\end{document}